\documentclass[5p,times,sort&compress]{elsarticle}
\usepackage[english]{babel}

\usepackage{amssymb}
\usepackage{amsmath}
\usepackage{xcolor}

\usepackage{microtype}

\newcommand{\dd}{\mathrm{d}}
\newcommand{\rmd}{\mathrm{d}}

\newcommand{\ie}{i.e.\ }

\journal{Computer Physics Communications}

\begin{document}

\begin{frontmatter}

\title{Hybrid deterministic/stochastic algorithm for large sets of rate equations}

\author[cea]{M.~Gherardi}

\author[cea]{T.~Jourdan\corref{cor1}}
\ead{thomas.jourdan@cea.fr}

\author[edf]{S.~Le Bourdiec}

\author[edf]{G.~Bencteux}

\cortext[cor1]{Corresponding author. Tel.:+33 1 69 08 73 44. Fax: +33 1 69 08 68 67}

\address[cea]{CEA, DEN, Service de Recherches de M\'etallurgie Physique, F-91191 Gif-sur-Yvette, France}
\address[edf]{EDF R\&D, F-92141 Clamart, France}

\begin{abstract}

  We propose a hybrid algorithm for the time integration of large sets
  of rate equations coupled by a relatively small number of degrees of
  freedom. A subset containing fast degrees of freedom evolves
  deterministically, while the rest of the variables evolves
  stochastically. The emphasis is put on the coupling between the two
  subsets, in order to achieve both accuracy and efficiency.
  The algorithm is tested on the problem of nucleation, growth and
  coarsening of clusters of defects in iron, treated by the formalism
  of cluster dynamics. We show that it is possible to obtain results
  indistinguishable from fully deterministic and fully stochastic
  calculations, while speeding up significantly the computations with
  respect to these two cases.  

\end{abstract}


\end{frontmatter}

\section{Introduction}

The use of rate equations covers a wide range of applications in
  fields such as chemistry, biology and physics. In material science
it has proved to be successful to study
long term evolution processes that affect the microstructure of
materials. This approach, also known as ``rate theory'' or ``cluster
dynamics'', enables one to study phenomena such as homogeneous
precipitation under thermal ageing~\cite{Jourdan2010} or formation of
voids and interstitial loops under
irradiation~\cite{Hayns1975,Ghoniem1979,Hardouin-Duparc2002}. The
system is considered as a gas of clusters, which are defined by their solute and self-defect
(vacancies, self-interstitials) content, and which can emit and absorb
mobile species. The evolution of each cluster type is given by a rate
equation, which also involves the number or
concentration of other types of clusters. 

In general, such a set of ordinary differential
  equations (ODE) is stiff, which roughly means that some modes have very
small time constants with respect to others. In such cases, implicit
integration methods are necessary, which implies the solving, at each time step, of a
linear system whose size is given by the number of cluster types considered in the simulation. Since it is
customary to simulate large clusters (containing several millions
single species), the solving of such an ODE set can become
challenging, from a point of view of both computation time and memory
storage. It is all the more the case when multi-component systems are
treated: besides industrial alloys, some simple cases include the
modelling of pure metals with solute impurities such as carbon,
nitrogen, etc., and gas atoms, such as helium and
hydrogen~\cite{ortiz2009}. When self-defects and a single type of
impurity are considered, some numerical approximations have proved
useful to reduce the number of
equations~\cite{Ghoniem1999,Golubov2007}. However, alternatives must
be envisaged once additional components are included in the
simulations.

One interesting way to solve these equations is to use a Monte Carlo
algorithm~\cite{Baskes1981}, as suggested by Gillespie in a general
framework~\cite{Gillespie1976}. At variance with other fields, where
stochastic effects must sometimes be taken into account to correctly
model noisy systems~\cite{Srivastava2002}, differences in the physical
results are not expected in our case, but using Monte Carlo
simulations leads to a tremendous decrease in memory storage. Since
each reaction firing has to be processed, it can however be
computationally demanding when reactions occur frequently; due to the
stiffness of the underlying ODE set, this case is quite common, and
approximations are needed. Some methods rely on the grouping of
events~\cite{Gillespie2001}, others are based on the partitioning
between slow and fast reactions, which are treated
differently~\cite{Haseltine2002}. Recently, a hybrid stochastic
and deterministic algorithm has been devised for small biochemical
systems, with a stochastically exact coupling~\cite{Alfonsi2005}. We
have adapted such an algorithm to the case of a large set of species,
coupled together by a small number of species. 

It should be emphasized here that the reason for the use of such an
hybrid method in material science is fundamentally different from what
motivates this kind of approach in biology. In general, for
biochemical systems, the deterministic equations are introduced as a
useful approximation of the chemical master
equation~\cite{Jahnke2011}, but can lead to erroneous results if the
approximation is not carefully controlled. In our case, we aim to speed up
simulation of systems which are already correctly described by
deterministic rate equations, by introduction of a stochastic part.

In the following we describe the algorithm in a general way and
present some simple test cases taken from material science, namely the
formation kinetics of voids and interstitial loops in iron and in iron
with helium under irradiation. Provided an approximation is made to
the initial algorithm devised for small biochemical
  systems, hybrid simulations are found to be significantly faster
than the purely deterministic and purely stochastic simulations.

\section{Background and notation}

We consider a well-stirred system of species (labeled by indices in a
finite set $\Lambda$) interacting through reaction channels (labeled
by indices in a finite set $\mathcal R$) in a volume $V$.  The
dynamical state of the system at time $t$ is specified by the set
$X_\Lambda(t)\equiv\left\{X_i(t)\right\}_{i\in\Lambda}$, whose
elements are the numbers of particles of type $i$.  The probability,
given the state at time $t$, that a reaction of type $j\in\mathcal R$
will occur in the time interval $[t,t+\rmd t]$ can be written under
general assumptions as \cite{Gillespie1992}
\begin{equation}
\label{eq:fundamentalassumption}
P\left(\mathrm{reaction}\;j\;\mathrm{during}\;[t,t+\rmd t]\;|\;X_\Lambda(t)\right)
=a_j\left(X_\Lambda(t)\right)\rmd t, 
\end{equation}
where $a_j$ is called the \emph{propensity function} for the $j$-th
reaction channel.  The change in the number of molecules of type $i$
produced by a reaction of type $j$ is called the \emph{state-change
  vector} $\nu_{ij}$.  All propensity functions and state-change
vectors completely specify the reaction channels.  Notice that
information about the space localization of the particles in the
physical system is completely neglected in this framework.

Eq.~(\ref{eq:fundamentalassumption}) lies at the heart of the
\emph{Stochastic Simulation Algorithm} (SSA), introduced by Gillespie
\cite{Gillespie1976}, which is based on the observation that the
probability at a given time $t_0$ that the next reaction is $j$ and
occurs in the time interval $[t,t+\rmd t]$ is $p(j,t|t_0)\rmd t$, with
\begin{equation}
\label{eq:nextreaction}
p\left(j,t|t_0\right) = a_j\left(X_\Lambda(t_0)\right)\;\exp\left[-(t-t_0) A_{\mathcal{R}}\left(X_\Lambda(t_0)\right)\right],
\end{equation}
where we have introduced the \emph{total propensity}
\begin{equation}
\label{eq:totalpropensity}
A_{\mathcal{R}}\left(X_\Lambda(t)\right) = \sum_{j\in\mathcal{R}} a_j\left(X_\Lambda(t)\right).
\end{equation}
Regardless of the reaction $j$, the probability that at a given time
$t_0$, a reaction occurs in $[t,t+\rmd t]$ is $p(t|t_0)\rmd t$, with
\begin{equation}
\label{eq:nextreaction-tot}
p\left(t|t_0\right) = A_{\mathcal{R}}\left(X_\Lambda(t_0)\right)\;\exp\left[-(t-t_0) A_{\mathcal{R}}\left(X_\Lambda(t_0)\right)\right].
\end{equation}
From Eq.~\eqref{eq:nextreaction}
and~\eqref{eq:nextreaction-tot}, one infers that the probability that
the reaction occurring in $[t,t+\rmd t]$ is of type $j$ reads
\begin{equation}
\label{eq:selection-typej}
\frac{p\left(j,t|t_0\right)}{p\left(t|t_0\right)} = \frac{a_j\left(X_\Lambda(t_0)\right)}{A_{\mathcal{R}}\left(X_\Lambda(t_0)\right)}.
\end{equation}

The evolution of the system is simulated by computing realizations of
the continuous-time Markov chain $X_\Lambda(t)$.  By calling $N_j(t)$
the stochastic process counting the number of times that a reaction of
type $j$ fires in the interval $[0,t]$ (which is an inhomogeneous
Poisson process of rate $a_j\left(X_\Lambda(t)\right)$ by Equation~(\ref{eq:fundamentalassumption})), one has
\begin{equation}
\rmd X_i(t)=\sum_{j\in\mathcal{R}}\nu_{ij}\rmd N_j(t).
\end{equation}
Operatively, at each time step $t_0$ of the SSA the following operations are performed:
\begin{enumerate}
\item Compute all propensities $a_j\left(X_\Lambda(t_0)\right)$, together with their sum $A_{\mathcal{R}}\left(X_\Lambda(t_0)\right)$.
\item Select a reaction $j$ with probability proportional to its propensity $a_j\left(X_\Lambda(t_0)\right)$, following Equation~\eqref{eq:selection-typej}.
\item Select a next-reaction time $t$ by sampling the probability distribution in 
Equation~(\ref{eq:nextreaction-tot}), which means that
  \begin{equation}
    \label{eq:choice-t-in-practice}
    t-t_0 = \frac{\xi}{A_{\mathcal{R}}\left(X_\Lambda(t_0)\right)},
  \end{equation}
  where $\xi$ is a random number exponentially distributed with parameter~1.
\item Update time ($t_0\mapsto t$) and configuration ($X_i\mapsto X_i+\nu_{ij}$ for each $i\in\Lambda$).
\end{enumerate}
A consequence of Equation~(\ref{eq:nextreaction-tot}) is that the larger
the total propensity $A_{\mathcal{R}}$ is, the slower the evolution.
In fact, the typical time between two reactions is
$1/A_{\mathcal{R}}$.  This problem is most painful when the system is
\emph{stiff} --- meaning that at least two very different time scales
are present --- and just a few reaction channels contribute most to
the total propensity.  In such a case the algorithm spends most of the
time performing fast reactions and taking small time steps.  Some
solutions to speed up the simulations have been proposed, based on
diverse ideas such as reuse of pseudo-random numbers
\cite{GibsonBruck2000}, quasi-steady-state theory \cite{RaoArkin2003},
grouping of reactions (the so-called \emph{tau leaping} methods)
\cite{Gillespie2001}.  In many applications where stochastic noise is
negligible for all practical purposes, such as in cluster dynamics
\cite{Hardouin-Duparc2002}, a commonly followed approach consists
instead in solving a deterministic system of equations describing the
evolution of average quantities.  More precisely, a well-known consequence of
Equation~(\ref{eq:fundamentalassumption}) is the \emph{chemical master
  equation} (see for instance \cite{Gillespie2000}), which in turn
implies the following \emph{rate equations}, provided fluctuations are neglected:
\begin{equation}
\label{eq:rateequations}
\frac{\rmd X_i(t)}{\rmd t}=\sum_{j\in\mathcal R}\nu_{ij}a_j\left(X_\Lambda(t)\right).
\end{equation}
In this equation, $X_i$ are now real functions of time. The precise
conditions under which such an evolution ``on average'' is a good
approximation of the original jump process are described in
Reference~\cite{Gillespie2000}.  In our case populations $X_i$ are
large for realistic volumes $V$, so we will assume that these
conditions are satisfied. The rate equations can then be integrated
deterministically using various numerical methods adapted to ODE
systems.

\section{The algorithm}

\subsection{The hybrid method}

The deterministic approach is not well-suited to situations where the
number of degrees of freedom is large, due to the poor scaling of
known numerical integration methods.  What we propose here is to adopt
a strategy similar to the one already proposed by Alfonsi \emph{et
  al.}~\cite{Alfonsi2005} for applications in biochemistry, namely
splitting the reactions into two sets, ideally according to their
typical time scale, and modeling one of them stochastically through
the SSA and the other one deterministically by using one among the
well-known numerical methods for ODE systems. In this subsection we
recall briefly the hybrid algorithm presented in Reference~\cite{Alfonsi2005}. 

Let the set of reactions $\mathcal{R}$ be partitioned into two disjoint subsets,
$\mathcal R =\mathcal D \cup \mathcal S$.
The evolution equation for $X_i(t)$ will be given by
\begin{equation}
\label{eq:hybridprocess}
\rmd X_i(t) = \sum_{j\in\mathcal{D}} \nu_{ij}a_j\left(X_{\Lambda}(t)\right)\rmd t +
\sum_{j\in\mathcal{S}}\nu_{ij}\rmd N_j(t),
\end{equation}
where $X_i(t)$ are real variables.
The idea is to evolve the system deterministically by using only the reactions in $\mathcal D$, 
and account for the second sum in the right-hand side by performing reactions 
in $\mathcal S$ stochastically.
More precisely, we consider the reduced system whose right-hand side function is given by the sum of the reaction terms on the set $\mathcal D$
\begin{equation}
\label{eq:reducedsystem}
\frac{\rmd X_i(t)}{\rmd t} = \sum_{j\in\mathcal D}\nu_{ij}a_j\left(X_\Lambda(t)\right),
\end{equation}
then at any given time $t_0$ we compute its numerical solution, with initial condition $X_i(t_0)$, up to the first-reaction time $t>t_0$, when we perform a stochastic reaction and update the configuration accordingly.
This stochastic reaction is instantaneous, in that it causes a change in the variables ($X_i
\mapsto X_i + \nu_{ij}$ for $i\in\Lambda$) without increasing time.

The problem of how to compute this reaction time is
non-trivial. Indeed, contrary to the standard SSA, propensities evolve
between two reactions treated stochastically due to the deterministic
evolution of the system. As shown by Alfonsi \emph{et
  al.}~\cite{Alfonsi2005}, it is possible to find the first-reaction
time by solving the following equation:
\begin{equation}
  \label{eq-solve-inhomogeneous}
  g(t|t_0) \equiv \int_{t_0}^t A_{\mathcal{S}}\left(X_\Lambda(\tau)\right)\,\rmd\tau = \xi,
\end{equation}
where $\xi$ is an exponentially distributed random number with parameter~1 and
$A_{\mathcal{S}}$ is the \emph{total stochastic propensity}
\begin{equation}
\label{eq:totalstochasticpropensity}
A_{\mathcal{S}}\left(X_\Lambda(t)\right)=\sum_{j\in\mathcal S} a_j\left(X_\Lambda(t)\right).
\end{equation}
One can see that Equation~\eqref{eq-solve-inhomogeneous} is a
straightforward generalization of
Equation~\eqref{eq:choice-t-in-practice} to the case when propensities
vary between two stochastic reactions.

\subsection{Partitioning the variables}

With the foregoing discussion, the original problem has been split
into a deterministic and a stochastic part, in such a way that the
latter is faster than a full SSA algorithm, provided that the sum of
all stochastic propensities (\ref{eq:totalstochasticpropensity}) is
less than the total propensity (\ref{eq:totalpropensity}).  On the
other hand, inspection of Equation~(\ref{eq:hybridprocess}) shows that
we have not gained much on the deterministic side, since the number of
equations to be treated has not changed.  In fact, reaction terms with
$j\in\mathcal D$ are still involved in the equations for all the
variables $X_i$ with $i\in\Lambda$.  In the type of hybridization
proposed in the biochemical literature one can afford having the
deterministic solver handle all species $X_\Lambda$ because one is
typically in the situation where the number of degrees of freedom is
modest and a full SSA would be slower.  In the present case instead we
want to take advantage of both approaches and tackle systems with a
large number of degrees of freedom.

In order to address the problem in a general way, we propose to split
the variables into two sets (in the same spirit as
in~\cite{Surh2008}), by identifying a proper subset $\Delta$ of
$\Lambda$ and treating variables with indices inside or outside
$\Delta$ differently.  A partition of the reactions is induced by the
choice of $\Delta$.
We impose that a reaction is treated deterministically if and only if
\emph{every} species involved in it (both the reactants and the
products) belongs to $\Delta$.  With
this definition we can rewrite Equation~(\ref{eq:hybridprocess}) as
\begin{equation}
\label{eq:partitionedprocess}
\left\{
\begin{aligned}
&\rmd X_{i\in\Delta}(t)=\sum_{j\in\mathcal D}\nu_{ij}a_j\left(X_\Delta(t)\right)\rmd t&+\sum_{j\in\mathcal S}\nu_{ij}\rmd N_j(t)\\
&\rmd X_{i\notin\Delta}(t)=&\sum_{j\in\mathcal S}\nu_{ij}\rmd N_j(t)
\end{aligned}
\right.
\end{equation}
where $X_\Delta(t)$ is a shorthand for the set of all $X_i$ with
$i\in\Delta$.  Species outside $\Delta$ evolve only through stochastic
terms.  Moreover, the deterministic reactions in the other set of
equations now only depend on the species inside $\Delta$ so that the
deterministic part of the method must deal with a reduced number of
degrees of freedom.  Namely, all one has to do in between two
successive stochastic reactions is to solve the upper-left corner of
the system (\ref{eq:partitionedprocess}), that is
\begin{equation}
\frac{\rmd}{\rmd t} X_{i\in\Delta}(t) = \sum_{j\in\mathcal{D}}\nu_{ij}a_j\left(X_\Delta(t)\right).
\end{equation}
Stochastic reactions are then triggered by looking at the time-change
function $g$ \eqref{eq-solve-inhomogeneous}.
General knowledge about deterministic solvers and the SSA, together
with the discussion at the beginning of this subsection, suggest
that we should choose $\Delta$ so that a few (fast) degrees of freedom
are treated deterministically, while the bulk of them should be left to the
stochastic part.

\subsection{Decoupling and the final algorithm}

Actual implementation of the hybrid method as it is described above
brings out some technical difficulties.  The main problem resides in
the abrupt changes in $X_i$ caused by the stochastic reactions, since
such variations in the right-hand side function can considerably slow
down the deterministic solving of the ODE set. In order to reduce the
severity of this behavior we introduce an approximation.

We define another subset $D$ of $\Lambda$, such that
$D\subseteq\Delta$, and decouple the deterministic and stochastic
dynamics inside $D$.  What this means is that whatever stochastic
reaction occurs, only the variables $X_i$ with $i\in\Delta\setminus D$
will be updated, while those inside $D$ will be left unchanged.  One
of course has to compensate in some way for the modification imposed
to the dynamics.  We do this by adding an effective source term
$\rho_i$ to each equation in $D$, crafted in such a way as to
approximately balance the discrepancy.  Namely, we take $\rho_i$ to be
the mean --- over all possible stochastic reactions $j$ --- of the
change that $X_i$ would have undergone if it were not decoupled,
weighted by the reaction's propensity:
\begin{equation}
\label{eq:effectivesource}
\rho_i = \sum_{j\in\mathcal S} \nu_{ij}a_j\left(X_\Lambda(t)\right).
\end{equation}
Since $X_i$ is not updated when a stochastic reaction occurs if $i\in D$, the deterministic solving is much more efficient, provided most stochastic reactions involving species in $\Delta$ actually involve species in $D$ and no species in $\Delta\setminus D$. A possible source of 
slowing down for the deterministic solver is the abrupt change of propensities in $\rho_i$ when a stochastic reaction occurs. However, in general the firing of a single stochastic reaction does not change $\rho_i$ appreciably, so the deterministic solver is virtually not affected.
We are not going to give analytical estimates of the errors this
approximation introduces to the dynamics; we justify it \emph{a
  posteriori} by checking the numerical results against exact
simulation methods.  Let us just note here that inclusion of such
effective source terms essentially restores the full right-hand side
of Equation~(\ref{eq:rateequations}) for the species in $D$.  

Finally, the complete algorithm is as follows:
\begin{enumerate}
\item \label{algorithm:xi}At time $t_0$ generate random number $\xi$
  exponentially distributed with parameter~1.
\item \label{algorithm:rho}Compute $\rho_i$ for each $i\in D$ as in (\ref{eq:effectivesource}).
\item Evolve the variables $X_i(t)$ with $i\in\Delta$ following
\begin{equation}
\label{eq:eqinD}
\left\{
\begin{aligned}
\frac{\rmd}{\rmd t} X_{i\in D}(t) &= \sum_{j\in\mathcal{D}}\nu_{ij}a_j\left(X_\Delta(t)\right) + \rho_i\\
\frac{\rmd}{\rmd t} X_{i\notin D}(t) &= \sum_{j\in\mathcal{D}}\nu_{ij}a_j\left(X_\Delta(t)\right).
\end{aligned}
\right.
\end{equation}
Keep track of the evolution of the total stochastic propensity
(\ref{eq:totalstochasticpropensity}) (which depends also on variables
in $\Delta$) and of the time-change function $g$ \eqref{eq-solve-inhomogeneous}.
\item \label{algorithm:stoc} When $g(t|t_0) = \xi$,
  choose a stochastic reaction $j\in\mathcal S$ with probability
  proportional to its propensity and update $X_i \mapsto X_i +
  \nu_{ij}$ for each $i\in\Lambda\setminus D$
\item Update $t_0$ and go back to step \ref{algorithm:xi}.
\end{enumerate}

\subsection{\label{section:implementation}Remarks on the implementation}

The method presented above leaves some open choices to the implementer.
Apart from the parameters $V$, $\Delta$ and $D$ (the tweaking of which should be done on a case-by-case basis; see Section~\ref{section:results}), some freedom is left as far as other details are concerned.

Let us consider the computation of the effective term $\rho_i$ at step \ref{algorithm:rho} of the algorithm.
One could choose not to compute it at every stochastic step, but rather to update it just once every $n_\rho$ reactions.
Moreover, one could choose to sample and hold its value between two updates, thus neglecting its smooth dependence on the variables inside the deterministic region $\Delta$ between two reactions.
We tested these different approximations for the cases described in Section~\ref{section:results} and found no big differences in performance or precision for values of $n_\rho$ between $1$ and $10^5$.

An important design-related issue concerns the data structure used to store the variables outside $\Delta$, which should be the large majority.
For the most frequently used reaction types \cite{Gillespie1977}, propensity functions are such that some of them vanish whenever a variable takes on the special value $0$.
The best choice is then to use a dynamical structure, so that slots are assigned to $X_{i\in\Lambda\setminus\Delta}$ only when it gets above zero, and they are released whenever it goes back to zero, in such a way as to take full advantage of the shape of the distribution $X_\Lambda$ and avoid the computation of useless propensities.
This is the implementation we use in the next section.

In order to solve the stiff set of ordinary differential equations in the
hybrid model (as well as in the fully deterministic simulations that
will be used as reference), a multistep method with backward
differentiation formulas  is used, as implemented in
CVODES from the SUNDIALS package \cite{Hindmarsh2005}.
This solver contains a root-finding procedure, which we use to detect time $t$
when $g(t|t_0)=\xi$, \ie when a stochastic reaction
occurs. Since the method is implicit, it requires the solving of a nonlinear system, which is done using
Newton iteration. The solving of linear systems in the Newton
iteration is carried out by a dense direct solver. Although more
sophisticated methods can be used for the solving of the linear
system, this one is chosen for its simplicity of implementation.

\section{\label{section:results}Results}

\subsection{Cluster dynamics model}

The model we use here, called cluster dynamics, describes the time
evolution of defect clusters in materials, within a mean-field
approach~\cite{Hardouin-Duparc2002}. Clusters (or species) can absorb
and emit some species that diffuse in an effective medium. Generally,
the set of  mobile species is limited to small clusters.

In the following we assume that clusters are made of two elements,
namely self-defects and solute (S) atoms, but the method could be extended
to more complicated clusters. Self-defects can either be vacancies (V) or
self-intertitial (I) atoms. Clusters are thus identified by $i=(n,p)\in
\mathbb{Z}\times\mathbb{Z}^+\setminus\{(0,0)\}$, where
$\left|n\right|$ is the number of self-defects ($n$ is positive for
self-interstitials and negative for vacancies) and $p\ge 0$ is the
number of solute atoms in the cluster. When no solute is present, clusters
are simply labeled by $i=n\in\mathbb{Z}\setminus\{0\}$.  Extended
defects, such as dislocations, grain boundaries and surfaces, are
treated as sinks and sources for mobile clusters. The evolution
equation for the concentration or number $X_i$ of species of type $i$
is
\begin{equation}
  \label{eq:evolution-eq-CD}
  \frac{\mathrm{d}}{\mathrm{d}t}X_i = G_i + \sum_{l \in M_i} J_{i-l,i} - \sum_{l \in M} J_{i,i+l} - \sum_{l \in \Lambda} J_{l,i+l} - k_i D_i (X_i - X^{e}_i).
\end{equation}
In Eq.~(\ref{eq:evolution-eq-CD}), $G_i$ is a creation rate of species $i$ by
irradiation, $k_i$ is the sink strength due to dislocations, grain
boundaries or surfaces, $D_i$ is the diffusion coefficient and
$X^{e}_i$ is the concentration or number of species $i$ in
thermodynamical equilibrium with the sinks (only valid for
self-defects). $\Lambda$ is the set containing all species and
$M$ contains only mobile species. The set
$M_i$ is a subset of $M$, to ensure that the
species involved in the flux $J_{i-l,i}$ exist (the number of
solute atoms of $i-l$ must be non-negative). The flux $J_{i,i+l}$
is defined by
\begin{equation}
  \label{eq:flux}
  J_{i,i+l} = \beta_{i,l} X_{i} X_{l} - \alpha_{i+l,l} X_{i+l},
\end{equation}
where $\beta_{i,l}$ and $\alpha_{i+l,l}$ are absorption and
emission rates respectively. The absorption rate is defined by solving
the diffusion equation in stationary state around an isolated cluster,
which leads to
\begin{equation}
  \label{eq:beta-CD}
  \beta_{i,l} = r_{i,l} D_{l}/V,
\end{equation}
where $r_{i,l}$ has the dimension of a distance and is related to the
geometry of clusters $i$ and $l$, and $V$ is the total simulation
volume if $X$ is a number of species, or is equal to 1 if $X$ is a
concentration. The emission coefficient is defined by
\begin{equation}
  \label{eq:alpha-CD}
  \alpha_{i+l,l} = \frac{r_{i,l} D_l}{V_{\mathrm{at}}} \exp{\left( -\frac{F_{i+l,l}}{\mathrm{k_B}T}\right)}.
\end{equation}

In this equation $V_{\mathrm{at}}$ is the atomic volume,
$F_{i+l,l}$ is the binding free energy of species $l$ to
cluster $i+l$, $\mathrm{k_B}$ is the Boltzmann constant and $T$ is the
temperature.

The number of ODEs to solve can amount to several
millions with only vacancies, self-interstitials and solute atoms. In
addition the propensities can be different by several orders of
magnitude. The resolution of such a stiff system must be done by an
implicit method, which involves the resolution of a linear system
based on the jacobian matrix. This can considerably slow down
the computations and even render the simulations impractical, mainly
for memory storage considerations.

However, the distribution of clusters is often sparse, which means
some concentrations can be considered as zero with a good
approximation. This is one of the motivations to use a stochastic
method: as only data related to clusters that exist in the simulation
box are stored in memory, and since the inversion of a linear system
is no longer necessary, we can expect to solve the memory
problems. Contrary to biochemical systems for example, where
stochastic methods have proved necessary to obtain reliable
results \cite{Srivastava2002}, we do not expect effects due to the
stochastic nature of the processes, mainly because the elementary
species that contribute most to the kinetics are present in large
numbers.

\begin{figure}[t]
\centering
\includegraphics[scale=1.3]{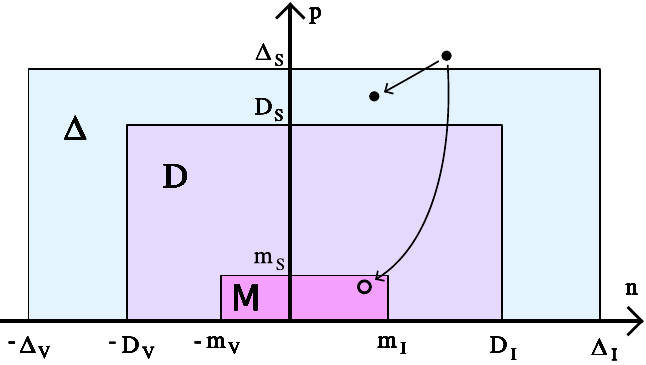}
\caption{(Color online) Partition of the set of clusters. 
Shaded areas correspond to the deterministic ($\Delta$), decoupling ($D$) and mobile ($M$) regions.
The arrows describe an emission reaction; the hollow point represents a cluster whose variable is not affected by the reaction, since it lies inside $D$.}
\label{fig:partitions}
\end{figure}

The equations of cluster dynamics (\ref{eq:evolution-eq-CD}) are
coupled to each other by mobile clusters, which are supposed to be
small clusters.  Each stochastic reaction will affect at least one
cluster in $M$, modifying the corresponding variable.  It is then
natural to choose the decoupling zone $D$ in such a way as to include
at least all these small and active clusters, in order to smooth out
part of their fluctuations.  Let $M$ be a rectangular region
$[-m_\mathrm{V},m_\mathrm{I}]\times[0,m_\mathrm{S}]$ as in
Figure~\ref{fig:partitions} (in general, if the set of mobile clusters
is bounded it suffices to take the smallest rectangle that contains
it).  Correspondingly, we define
$D=[-D_\mathrm{V},D_\mathrm{I}]\times[0,D_\mathrm{S}]$ for the
decoupling region and
$\Delta=[-\Delta_\mathrm{V},\Delta_\mathrm{I}]\times[0,\Delta_\mathrm{S}]$
for the deterministic one (all these sets do not contain $(0,0)$, but
we will omit this in the notation for clarity).  Consider a non-mobile
cluster made of $n>m_\mathrm{I}$ interstitial atoms and
$p>m_\mathrm{S}$ solute atoms, corresponding to $X_{(n,p)}$.  If
$n\le\Delta_\mathrm{I}- \max(m_\mathrm{V},m_\mathrm{I})$ and
$p\le\Delta_\mathrm{S}-m_\mathrm{S}$ then no stochastic reaction can
affect $X_{(n,p)}$ (the same reasoning holds symmetrically for
vacancy-type clusters).  Therefore, any choice of $D$ such that
$m_\mathrm{I}<D_\mathrm{I}\le\Delta_\mathrm{I}-\max(m_\mathrm{V},m_\mathrm{I})$,
$m_\mathrm{V}<D_\mathrm{V}\le\Delta_\mathrm{V}-\max(m_\mathrm{V},m_\mathrm{I})$
and $m_\mathrm{S}<D_S\le\Delta_\mathrm{S}$ will have the same effect
on the algorithm. To highlight the advantages of such a choice, it is
useful to specify more precisely the rate equations for clusters
in $D$ under the same form as Eq.~\eqref{eq:eqinD}:
\begin{align}
  \frac{\dd }{\dd t} X_i &= \sum_{j\in\mathcal{D}}\nu_{ij}a_j\left(X_\Delta(t)\right) + \rho_i^{+} && i \in D \setminus M   \label{eq:evol-CD-DmM} \\
  \frac{\dd }{\dd t} X_i &= \sum_{j\in\mathcal{D}}\nu_{ij}a_j\left(X_\Delta(t)\right) + \rho_i^{+} - \rho_i^{-} X_i && i \in  M.\label{eq:evol-CD-M}
\end{align}
For $i\in D \setminus M$, $\rho_i^+$ is simply equal to $G_i$ ;
although in previous sections, this term was formally included in the
reaction terms, we write it here separately for more clarity,
restricting reaction terms to fluxes between clusters. 
It should be noted that, as a consequence of the choice of $D$,
Eq.~\eqref{eq:evol-CD-DmM} is the full evolution equation for
clusters in $D\setminus M$ and requires no additional stochastic contribution.
This is not the case for $i\in M$, where stochastic contributions are in
general very large since these clusters are coupled to all existing
clusters. However, these stochastic contributions are all contained in
the source term $\rho_i$, which can be written in terms of two quantities:
\begin{align}
  \rho_i^+ &= G_i + \sum_l \alpha_{l+i,i} X_{l+i} + k_i D_i X_i^e \label{rhoip} \\
  \rho_i^- &=  \sum_l \beta_{l,i} X_l + k_i D_i, \label{rhoim}
\end{align}
where the summation is performed on all indices that are not taken into
account through reactions in $\mathcal{D}$, \ie on $l$ such that $l
\in \Lambda \setminus \Delta$ and/or $l +i \in \Lambda \setminus
\Delta$. In practice, these source terms are updated every $n_\rho= 1000$
reactions. In order to avoid the entire recomputation of these source
terms, it is useful to keep track of the changes induced by the
stochastic reactions by updating a copy of $\rho_i^+$
and $\rho_i^-$ at every stochastic reaction.

\subsection{Iron under irradiation}

As a first application we consider the cluster dynamics description of
pure $\alpha$-iron under irradiation by neutrons.  The set $\Lambda$
labeling the species is the set of non-zero integer numbers (no solute
is present), and the deterministic and decoupling regions are
punctured intervals in $\mathbb{Z}\setminus\{0\}$.  The dynamical
variables $X_n$ will represent the number of clusters of type $n$, but
will nonetheless be treated as continuous quantities in the
deterministic sector~$\Delta$.  Vacancy clusters are mobile up to size
$m_\mathrm{V}=4$ and interstitial clusters up to $m_\mathrm{I}=3$.
The creation rate is adapted to describe cascades and is non zero for
monomers ($n=\pm 1$), clusters of $4$ interstitial atoms and of
$8$~vacancies.  The total irradiation time reached is around
$10^6\,\mathrm{s}$ ($\approx 11.6$ days) and the damage rate is
$3.9\times 10^{-8}$~dpa/s (dpa: displacements per atom). Other
parameters of the model are taken from Reference~\cite{Meslin2008}.

The choice of $\Delta$ is found to have no influence
  on the physical result, so it should only be driven by performance
  considerations. A minimal requirement for $\Delta$ is that it
should include all species produced by cascades, since irradiation is
a potentially fast reaction channel.  Apart from that, the choice of
$\Delta_\mathrm{V}$ and $\Delta_\mathrm{I}$ has to be adapted to the
characteristics of the system. In the $\alpha$-iron case we expect
that the distribution of $X_n$ becomes hollow in the vacancy sector at
large times, at least for sufficiently high temperatures due to
thermal dissociation of clusters, with one peak at small sizes (due to
the creation by irradiation) and one at large negative values of $n$.
The best choice for $\Delta_\mathrm{V}$ then probably lies in the
middle.  In the following we will report performance
results for two different choices of the $\Delta$ region.

\begin{figure}[t]
\centering
\hspace{-0.5cm}
\includegraphics[scale=0.95]{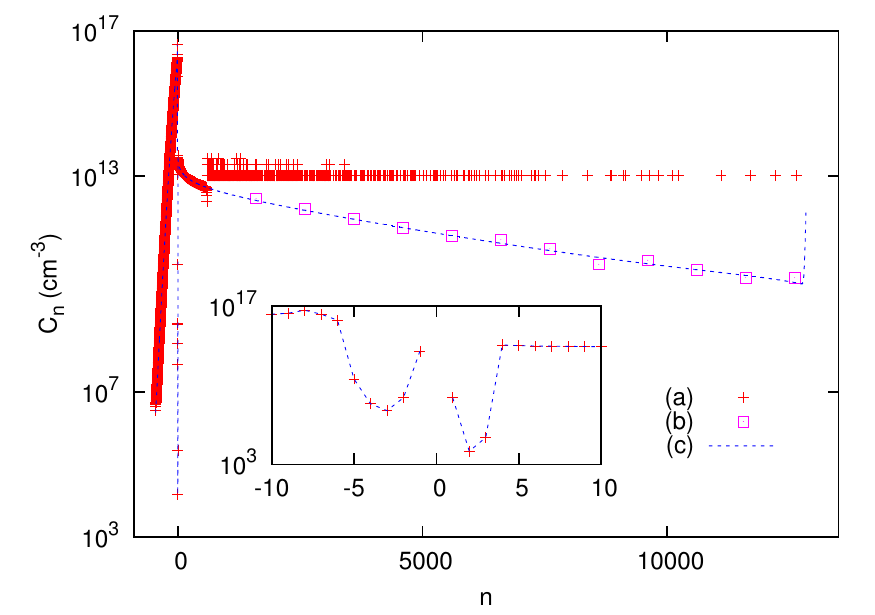}
\caption{(Color online) Cluster distribution at $t=4\cdot
  10^5\,\mathrm{s}$ for pure iron at $300^\circ\mathrm{C}$ (simulation
  volume is $V=10^{-13}\,\mathrm{cm}^3$).  Red crosses (a) correspond
  to the hybrid algorithm; pink squares (b) are histograms of width
  1000, obtained from the same data-set; the dashed blue line (c) is
  obtained by a purely deterministic method.  The inset shows a
  close-up for small clusters.}
\label{fig:fedistributions}
\end{figure}

Figure~\ref{fig:fedistributions} shows the distribution of clusters
$C_n\equiv X_n/V$ at $t=4\cdot 10^5\,\mathrm{s}$ and
$T = 300^\circ\mathrm{C}$. Results obtained both by our hybrid algorithm
in a single run and by a purely deterministic method are presented.
The simulation volume is $V=10^{-13}\,\mathrm{cm^3}$; clearly, all
points in the stochastic region outside $\Delta$ lie on or above the
threshold $1/V$.  In order to obtain points more easily comparable
with the exact solution, we consider binned data (see figure caption).
The shape of the distribution is correctly reproduced, both in the
stochastic and deterministic regions, the sole deviations being close
to the interface.  The distribution of small clusters is
indistinguishable from the expected one on the scale of the figure:
the effective corrections (\ref{eq:effectivesource}) are successful,
even if their values are not updated at every iteration.

\begin{figure}[t]
\centering
\hspace{-0.5cm}
\includegraphics[scale=0.95]{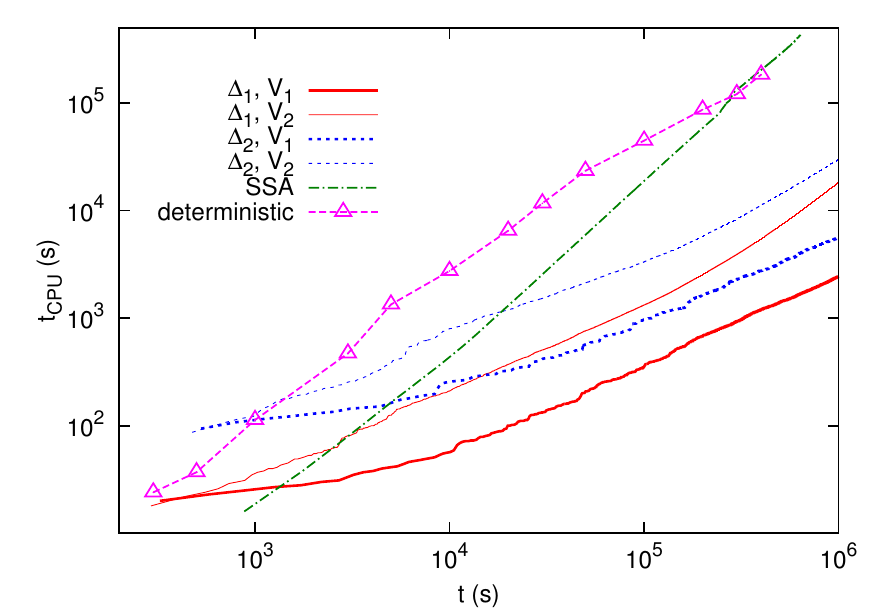}
\caption{(Color online) CPU time versus physical time (pure iron at
  $300^\circ\mathrm{C}$) for different methods and several choices of
  the parameters ($\Delta_1=[-300,300]$, $\Delta_2=[-450,600]$,
  $V_1=10^{-14}\,\mathrm{cm}^3$, $V_2=10^{-13}\,\mathrm{cm}^3$).  Pink
  triangles correspond to the purely deterministic method; the
  dot-dashed green line to the SSA; the remaining four lines are
  obtained by the hybrid algorithm, the two dashed blue lines
  corresponding to the wider $\Delta$ and the two solid red lines to
  the narrower.}
\label{fig:feperformance}
\end{figure}

\begin{figure}
\centering
\hspace{-0.5cm}
\includegraphics[scale=0.95]{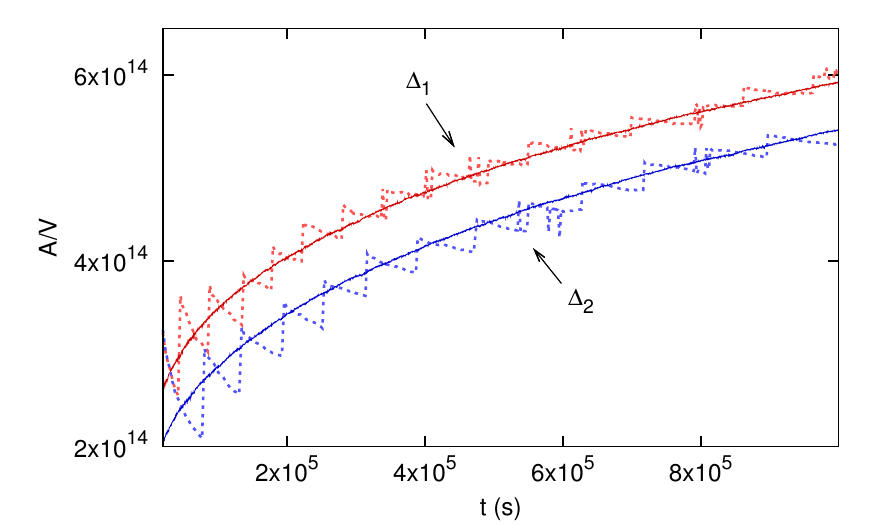}
\caption{(Color online) Evolution of the total stochastic propensity
  (in $\mathrm{s}^{-1}$) normalized by the volume (in $\mathrm{cm}^3$)
  for pure iron at $300^\circ\mathrm{C}$.  The two red curves on top
  correspond to $\Delta_1=[-300,300]$, the two blue curves underneath
  to $\Delta_2=[-450,600]$.  In each couple, the fluctuating dashed
  curves correspond to $V=10^{-15}\,\mathrm{cm}^3$ and the smoother
  ones to $V=10^{-13}\,\mathrm{cm}^3$ (The data for
  $V=10^{-14}\,\mathrm{cm}^3$ is not shown here for clarity, since
  points collapse very tightly onto the curves corresponding to the
  higher volume).}
\label{fig:fepropensity}
\end{figure}

Figure~\ref{fig:feperformance} shows the CPU time as a function of the
physical time reached (simulations were carried out on an Intel Xeon
X5650 2.66 GHz).  We report the performances for a purely
deterministic method, a standard SSA algorithm and our hybrid
algorithm. The SSA simulation was done with a volume
$V=10^{-13}\,\mathrm{cm^3}$. The performances obtained by the hybrid
algorithm are shown for two different simulation volumes
($V=10^{-14}\,\mathrm{cm^3}$, $10^{-13}\,\mathrm{cm^3}$) and two
choices of the deterministic interval ($\Delta=[-300,300]$,
$[-450,600]$).  The deterministic curve was obtained by first running
a hybrid simulation up to $t=10^6\,\mathrm{s}$ and noting the size of
the largest populated interstitial and vacancy clusters (let us call
them $I(t)$ and $V(t)$ respectively) at several intermediate times
$t_i$.  Afterwards, purely deterministic solutions were calculated at
said times, for the system constituted of the $N(t_i)= I(t_i)+V(t_i)$
equations describing interstitial clusters not larger than $I(t_i)$ and
vacancy clusters not larger than $V(t_i)$.  One of the advantages of our
hybrid approach is that the number of stochastic species is not fixed
before the simulation, and the number of degrees of freedom involved
in the computation automatically follows the growing number of
populated species.  As the plot reveals, the hybrid algorithm
outperforms both the purely deterministic solving and the standard SSA
solving, and has the additional advantage of being highly tunable.

Changing $V$ and $\Delta$ has a big impact on the algorithm's
behavior.  Increasing $V$ has the direct consequence of decreasing the
threshold $1/V$ and thus increasing the number of clusters
participating in stochastic reactions: the total stochastic propensity
$A_{\mathcal{S}}$ increases approximately linearly with $V$ (see
Figure~\ref{fig:fepropensity}); as a consequence, the CPU time it
takes to reach a certain physical time increases with the volume.
Decreasing the width of the deterministic interval
$\Delta$ has a positive effect in the short term, since it decreases
the strain on the deterministic solver, but it affects negatively the
long-time behavior, for the total propensity stays larger.  The mutual
influence of $\Delta$ and $V$ on the performance is a model-dependent
aspect that has to be considered carefully when trying to find the
best trade-off between speed and precision.

It is important to note that coping with the growth of $N(t)$ is the
main weakness of purely deterministic methods.  If the number of
equations is fixed, then a deterministic solution of the equations is
asymptotically faster than the SSA.  Indeed, for our problems,
increasingly large time steps are taken by the solver in the fully
deterministic scheme, whereas in the hybrid one the time step is
constrained by the root finding procedure, \ie by the firing of
stochastic reactions.  However, since in the Newton iteration the
matrix to be inverted is considered as dense and a direct linear
solver is used, the CPU time scales as $N^3$, and the memory storage
as $N^2$. The memory storage issue can even render calculations
unfeasible. In the hybrid case, $N$ should be replaced by the size of
the deterministic region $(\Delta_{\mathrm{I}}+\Delta_{\mathrm{V}}+1)\times (\Delta_{\mathrm{S}}+1)$,
which is far smaller than $N^2$. This is why if $N(t)$ is an
increasing function of time, the hybrid algorithm becomes better in
the long run from the point of view of CPU time and memory storage.
In the present case, $N(t)$ grows sublinearly in $t$.  A fit of the
form $N(t)\propto t^\alpha$ gives $\alpha\approx 0.35$.  Systems where
$N(t)$ grows faster than this will benefit even more from the hybrid
approach (see Section \ref{section:fehe}).

\begin{figure}[t]
\centering
\hspace{-0.5cm}
\includegraphics[scale=0.95]{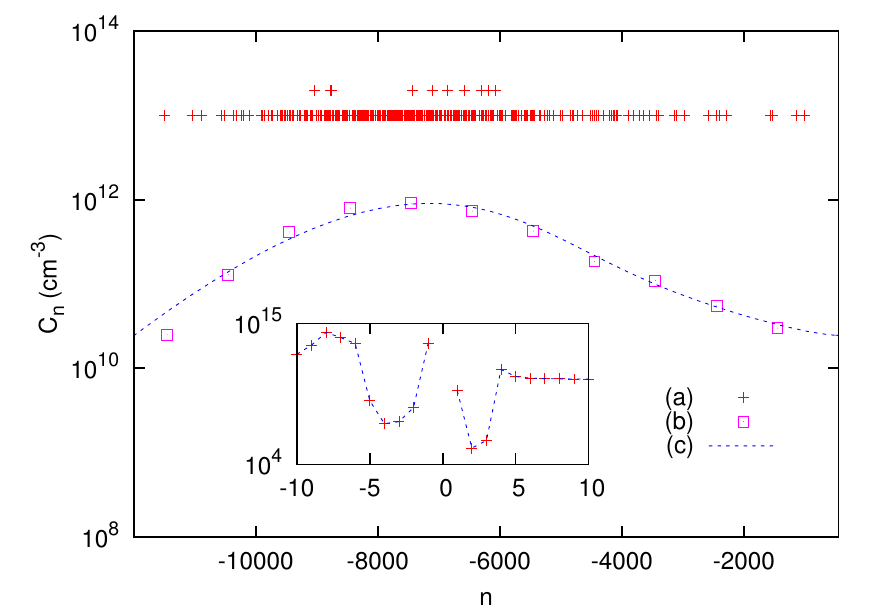}
\caption{(Color online) Cluster distribution at $t=4\cdot 10^5\,\mathrm{s}$ for pure iron at $400^\circ\mathrm{C}$
(simulation volume is $V=10^{-13}\,\mathrm{cm}^3$).
See the caption to Figure~\ref{fig:fedistributions} for an explanation of the symbols.}
\label{fig:fedistributions2}
\end{figure}

As we already mentioned, raising the temperature yields peaked
distributions. The dynamic implementation of the stochastic part of
our algorithm (see Section \ref{section:implementation}) permits to
take advantage of these situations by neglecting the degrees of
freedom corresponding to the hollow regions.  An example of such a case is shown in 
Figure~\ref{fig:fedistributions2}, where a second peak
is present, corresponding to very large vacancy-type clusters.  Again,
the position and shape of the peak (as well as the distribution at
small sizes) are well reproduced.

\subsection{\label{section:fehe}Iron/helium under irradiation}

The case of helium in iron is a particulary interesting one, from both
physical and numerical point of views. From the physical side, helium
is created by transmutation during irradiation with neutrons. As a
noble gas, its interaction with iron atoms is repulsive and it thus
tends to agglomerate in vacancies and vacancy clusters, forming helium
bubbles~\cite{Trinkaus2003}. Due to the helium pressure inside these
bubbles, the emission rate of vacancies is reduced as the
helium-to-vacancy ratio increases. At high temperature, only bubbles
containing enough helium remain stable, which can lead to sparsely
populated distributions. This is a case where the SSA and hybrid
simulations should perform particularly well. Very recently, SSA
calculations have been performed in iron containing both helium and
hydrogen~\cite{Marian2011}.

For clusters containing no helium, the parametrization is the same as
in the previous section, except the source term which corresponds to
ion irradiation~\cite{Meslin2008}. The damage rate is $10^{-4}$~dpa/s
and the helium-to-dpa ratio is $10^{-4}$. We intend to model
irradiation conditions close to those obtained by dual ion beam
experiments, where one beam is used to create damage (vacancies and
self-interstitials), while the other one injects helium into the
sample. Compared to irradiations performed with neutrons, where helium
is created by transmutations, the helium content can be significantly
higher in dual beam irradiations, so a role of helium can be more
clearly highlighted.

Other input data of our model, specific to helium, include the
diffusion coefficient of helium, deduced from first principle
calculations~\cite{Fu2005b}, and binding energies of helium atoms,
vacancies and interstitials to bubbles. These data are taken from
first principle calculations for small clusters~\cite{Fu2007} and
extrapolated from molecular dynamics calculations for larger
sizes~\cite{Lucas2009}. For the sake of simplicity, interstitial
clusters are assumed to contain no helium atoms.  In the following we
fix the volume to
$V=10^{-13}\,\mathrm{cm}^3$ and the
deterministic region to
$\Delta=[-10,10]\times[0,2]$. The
temperature is set to $500^\circ\mathrm{C}$.

Adding a solute atom to the picture heavily affects the computational
cost of a completely deterministic approach, since the space of
cluster types is now two-dimensional.  This has an impact both on the
speed of the simulations and on the memory requirements.  The latter
are especially important in the present case, where 1.3 GB of RAM are
needed in order to reach $t=3\,\mathrm{s}$ in a fully deterministic
computation.  Clearly, reaching much longer times with such an
approach is unfeasible.  On the other hand, the memory requirements of
the hybrid algorithm are much less demanding (less than 10 MB for all
the times considered), since the deterministic region is fixed and the
number of variables in the stochastic region follows the shape of the
distribution, as explained above.

\begin{figure}
\centering
\includegraphics[scale=1.07]{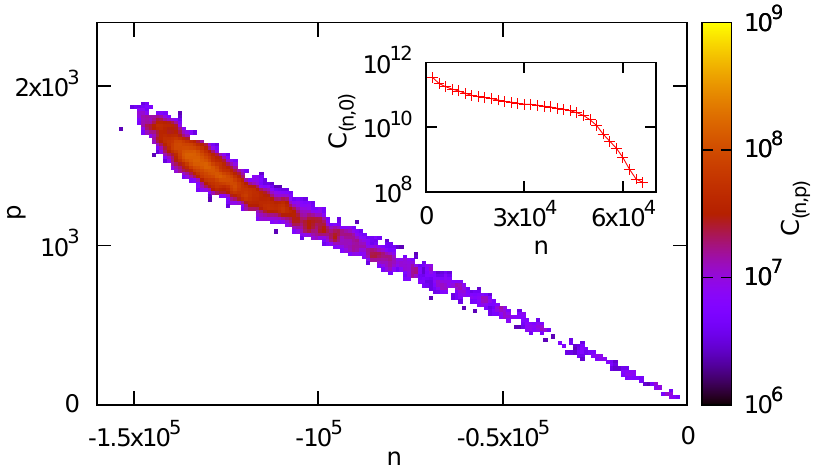}
\caption{(Color online) Distribution of clusters at
  $t=1000\,\mathrm{s}$ obtained by the hybrid algorithm in the
  iron/helium case, at $500^\circ\mathrm{C}$.  Data have been averaged
  on 50 simulations and smoothed by taking histograms: each point in
  the mesh represents the average on a region of size $1000\times 20$
  around it.  The inset shows the distribution of interstitial
  clusters, containing no helium, obtained by taking histograms of
  size 2000.  Concentrations $C_{(n,p)}$ are in $\mathrm{cm}^{-3}$.  }
\label{fig:fehedistributions}
\end{figure}

This distribution is shown at $t = 1000\,\mathrm{s}$ in
Fig.~\ref{fig:fehedistributions}, where data have been averaged on 50
simulations and grouped into histograms. Only a small part of the set
of clusters is populated, which supports the use of a dynamical
structure to store concentrations, neglecting non-populated
classes. If such an approach were not considered, the memory needed
at $t = 1000\,\mathrm{s}$ would amount to around 3 GB instead of the
few MB previously mentioned.

Concerning the computation time, the hybrid method is shown to be
faster than standard SSA by more than one order of magnitude for small
times and roughly three times faster at $t=500~\mathrm{s}$. Deterministic
calculations could only be performed until 3~s of physical time due to
memory requirements and are considerably slower than SSA and hybrid
calculations. The number of equations $N(t)$ corresponding to all
species inside the rectangle containing the set of clusters grows as
$N(t)\sim t^\alpha$, with $\alpha = 1.6$, which is much faster than the
strongly sublinear growth observed for pure iron. This is why the
deterministic solving fails so early. However, it should be noted that
using a dense solver is clearly not optimal in the present case, since
the jacobian matrix is sparse. Using a sparse direct solver, a
deterministic algorithm could prove to be convenient.
 Since any improvement on the linear solver will be
reflected in the hybrid algorithm, in general the hybrid algorithm
should perform at least as well as the deterministic one, while
possibly relieving technical difficulties associated with the solving
of large matrices.

\begin{figure}
\centering
\hspace{-0.5cm}
\includegraphics[scale=0.95]{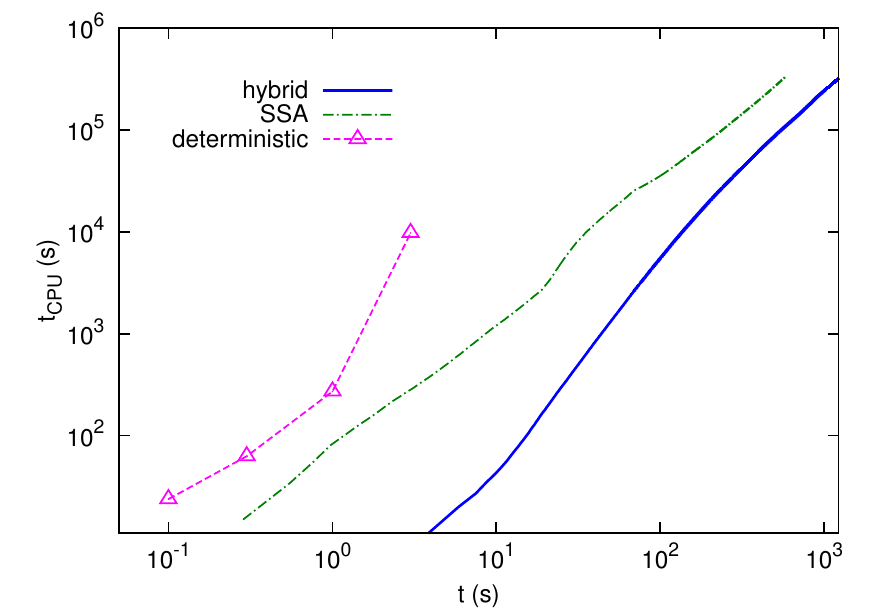}
\caption{(Color online) CPU time versus physical time (iron/helium case).
Pink triangles correspond to the purely deterministic method,
the dashed green line to the purely stochastic (SSA) method, and
the blue solid line to the hybrid algorithm.}
\label{fig:feheperformance}
\end{figure}

\section{Conclusions}

We proposed a new hybrid method for dealing with
large sets of stiff rate
equations, coupled by a small set of degrees of freedom, as those
appearing in cluster dynamics. In this approach deterministic and stochastic strategies work in
parallel across the time evolution.  Essentially, a fixed small number
of fast degrees of freedom is treated deterministically, while a
non-constant (and possibly very high) number of the remaining
variables is treated stochastically.  The coupling between the two
sectors is realized by the introduction of effective source terms for
a small number of equations, where fluctuations due to the stochastic
reactions are averaged out.  This approximation does not yield
appreciable errors in the results, and is helpful in reducing the time
needed for the simulations.

We tested the hybrid method in the cluster dynamics description of two
physical phenomena: the formation of voids and interstitial loops in
pure iron under irradiation, and the nucleation and
  growth of helium bubbles in the same system.
In the two
cases, a significant decrease in computation time is observed with
respect to both deterministic solving and standard SSA
algorithm. Since the number of variables treated deterministically in
the hybrid algorithm is far lower than in a fully deterministic
solving, the memory storage is also orders of magnitude lower, and
comparable to SSA. We emphasize that we have used a dense solver for
the deterministic and hybrid methods; using a sparse solver would
improve the performance of these two methods with respect to SSA. In
particular, a larger number of deterministic variables could be used
in the hybrid method, to keep a balance between the solving of the
deterministic and stochastic parts.

\section*{Acknowledgments}
\label{sec:acknowledgments}

This work was partially funded by the French National Research Agency
(ANR) through the ParMat Projet ANR-06-CIS6-006-03. We are much
indebted to V.~V.~Bulatov for suggesting the use of SSA to solve the
physical problems mentioned in the present article.

\end{document}